# The CHRONIOUS Ontology-Driven Search Tool: Enabling Access to Focused and Up-to-Date Healthcare Literature

Stephan KIEFER[1], Jochen RAUCH[1], Riccardo ALBERTONI[2],
Marco ATTENE[2], Franca GIANNINI[2], Simone MARINI[2],
Luc SCHNEIDER[3], Carlos MESQUITA[4], Xin XING[5], Michael LAWO[5]

[1]Home Care / Telemedicine- Fraunhofer Institute for Biomedical Engineering, St.Ingbert Germany
Tel. +49 (0)6894 980 (156) (215) Fax: +49 (0)6894 980400, Email: name.surname@ibmt.fraunhofer.de
[2]Institute for Applied Mathematics and Information Technology – C.N.R., 16149 Genova, Italy
Tel. +39 010 6475 (697)(691)(666)(696), Fax: +39 010 6475660 Email: name.surname@ge.imati.cnr.it
[3]Institute for Formal Ontology and Medical Information Science,
Saarland University, P.O. Box151150, Saarbrücken, 66041, Germany
Tel: +49-(0)681-302 64 770, Fax: +49-(0)681-302 64 772, Email: luc.schneider@ifomis.uni-saarland.de
[4]Link Consulting, Lisbon, 1000-138, Portugal,
Tel. +351 213 100 031, Fax: +351 213 100 008  Email: carlos.mesquita@link.pt
[5]TZI - Center for Computing and Communication Technologies, Am Fallturm 1, Bremen, 28359, Germany
Tel: +49 421 2183(035)/(824), Fax: +49 421 2187196, Email: surname@tzi.de

**Abstract:** This paper presents an advanced search engine prototype for bibliography retrieval developed within the CHRONIOUS European IP project of the seventh Framework Program (FP7). This search engine is specifically targeted to clinicians and healthcare practitioners searching for documents related to Chronic Obstructive Pulmonary Disease (COPD) and Chronic Kidney Disease (CKD). To this aim, the presented tool exploits two pathology-specific ontologies that allow focused document indexing and retrieval. These ontologies have been developed on the top of the Middle Layer Ontology for Clinical Care (MLOCC), which provides a link with the Basic Formal Ontology, a foundational ontology used in the Open Biological and Biomedical Ontologies (OBO) Foundry. In addition link with the terms of the MeSH (Medical Subject Heading) thesaurus has been provided to guarantee the coverage with the general certified medical terms and multilingual capabilities.

## 1. Introduction

Chronic diseases are mostly characterized by complex causality, multiple risk factors, long latency periods, prolonged course of illness and functional impairment or disability.

Chronic diseases may get worse, lead to death, be cured, remain dormant or require continual monitoring. It is clear that reducing the severity of both the symptoms and impact would mean significant benefit for both the individual and the society. Many physicians recognize that the long-term health outlook for chronic disease has improved in the past decade. Most of them credit the improvement to better management and monitoring.

Unforeseen possibilities for the continuous remote monitoring of patients are arising from modern sensors for the acquisition of environmental, biological and physical signals. This is turning out in novel knowledge pertaining to previously unconsidered signals, symptoms, commonalities which are pivotal for determining novel and more targeted care plans. Exploiting such new knowledge while searching and retrieving medical literature is therefore of fundamental importance for healthcare professionals.

To support the remote management of persons at risk or with chronic health conditions, the EU IP project CHRONIOUS (An Open, Ubiquitous and Adaptive Chronic Disease Management Platform for COPD and Renal Insufficiency) is developing an open framework for their monitoring and treatment. Chronic Obstructive Pulmonary Disease (COPD) and Chronic Kidney Disease (CKD) are two case studies to set up a test bed in the project, but the CHRONIOUS architecture is suitable for any chronic disease.

To deal with the continuous evolving literature in the COPD and CKD domains, the CHRONIOUS framework includes an intelligent literature Search Module that will be described in this paper. The CHRONIOUS Search Module enables healthcare professionals to access well-focused and up to date healthcare information in the form of scientific literature, guidelines, hospital-specific documentation and grey literature.

## 2. Adopted Methodology

The Search Module combines domain ontologies (one for each chronic disease treated) with the linguistic capabilities provided by the Medical Subject Heading thesaurus (MeSH) [1]. The domain ontologies are built on top of the Middle Layer Ontology for Clinical Care (MLOCC) [2], which guarantees the exploitation of well-funded and formalized medical concepts and the integration with other medical domain ontologies (e.g., ACGT Master Ontology [3], Open Biological and Biomedical Ontologies (OBO) foundry [4,5]). Based on medical guidelines, two domain-specific ontologies have been defined for the two case studies (COPD and CKD) of the project, and properly validated by clinical experts coordinated by an international medical board. Within CHRONIOUS, ontologies and MeSH provide complementary benefits; for each specific disease, indeed, an ontology provides a fine-grained knowledge that MeSH is not supposed to reach; on the other hand, MeSH provides well-established generic medical terminology, multilingual representations, synonyms, narrower/broader/related concepts which can be exploited by the clinicians for the composition of expressive queries in a simple and effective way. The combination of the domain specific ontologies and the MeSH thesaurus allows CHRONIOUS to provide more focused results when compared to traditional search engines for the medical literature. Moreover, to deal with the problem of medical knowledge evolution, which is even more evident in remote patient monitoring being at its early studies, the developed system provides tools for supporting the experts in upgrading the provided ontologies according to the new emerging concepts in the concerned documentation.

## 3. The Devised Search System Architecture

The architecture of the CHRONIOUS Search Module, which is depicted in Figure 1, relies on different sub-modules. Upload Tools are used to load documents related to CKD and COPD in the CHONIOUS System for their indexing. A Transformation Module converts the imported documents to normalized plain text to be processed by the Natural Language Processing (NLP) Tool, which allows to specify processing pipelines (e.g. consisting of Sentence Splitter, Tokeniser, Part-of-speech Tagger and Morphological Analyser) as well as filter parameters (e.g. for word kinds and categories) to extract the root words in the documents for the Indexer. In addition, it offers pre-linguistic processing services to the Ontology Enrichment tool in order to extract candidate concepts from the documents. The Indexer/Concept Associator generates the indices of each document imported into the system and associates concepts of the Ontology/Thesaurus with the concepts found in the documents. A Knowledge Cache maintains a graph database to store the generated indices and connections (Ontology, Thesaurus, Documents). The Search Module offers search methods as well as several tools that facilitate the user's interaction with the system

  

including the ontology/thesaurus browsing and the query building functionalities. The most interesting modules are deepened in the following sections.

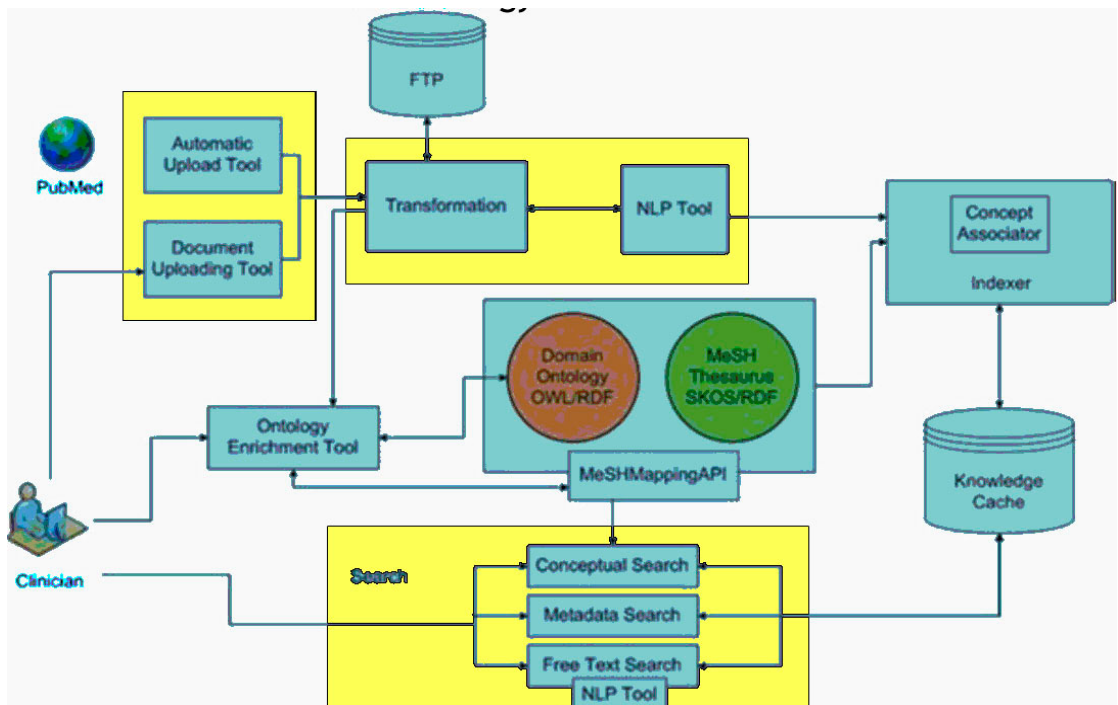

*Figure 1 : CHRONIOUS Ontology Module Architecture*

## 4.   Natural Language Processing

NLP techniques can be used for Information Extraction (IE), whose goal is to automatically extract structured information (i.e. categorized and contextually and semantically well-defined data from a certain domain) from unstructured machine-readable documents. In CHRONIOUS the NLP Tool processes the normalized document contents obtained from the Feed and Transformation Module and provides the specific annotations to the Document Indexer and Concept Associator as well as to the Ontology Enrichment Tool.

CHRONIOUS NLP is based on GATE [6], a leading resource for developing and deploying software components that process human language. GATE provides a framework or Java class library that can be used to embed language processing capabilities in diverse applications. The NLP Tool realizes the definition of the following default processing pipeline to extract the headwords of a text: segments text into sentences; splits the text into very simple tokens such as numbers, punctuation and words of different types; produces POS tags (e.g. noun, proper noun, preposition) as an annotation on each word or symbol; identifies the lemma (headword) and affix (e.g. "runs", "ran" and "running" has all "run" as lemma) by considering one token and its POS tag, one at a time.

For pre-linguistic processing requirements of the Ontology Enrichment Tool and the Indexer, the following resources have been further integrated in the NLP Tool:

- OntoRoot Gazetteer [7]: A GATE plugin to produce ontology-aware annotations for extracted terms.
- Shallow Parser: Analyses the sentences to identify word groups by linguistic patterns (e.g. "chronic disease", "lung function"). For the prototype implementation (English language), we extracted and utilized a JAPE transducer from the Text2Onto project [8].
- RegEx-Pattern Matcher: Matches the lemma of a token with word patterns defined as regular expressions.

 

- Dictionary Matcher: Matches the lemma of a token to a (common) dictionary. The JAPE resource, which we have implemented for English documents, uses the WordNet dictionary [9] and Java API for searching.
- Thesaurus Matcher: Matches the lemma of a token to a (domain) thesaurus. For the target medical domain we have implemented a JAPE resource, which uses the MeSH-Mapping API to access MeSH terms and the mapped ontology concepts.

The NLP Tool runs on a server and can be accessed through a web service that represents the communication layer, and provides methods for receiving documents in different formats and responds with processed data.

## 5. Search Module

This module offers three different search methods as well as several tools that facilitate the user's interaction with the system and with the ontology/thesaurus browsing and the query building functionalities, in particular:

- Metadata Search: a keyword-based search that looks specifically into the metadata associated with a document and indexed in the Knowledge Cache.
- Conceptual Search: the user submits queries at a higher level of abstraction (as compared to keyword-based search) that can be formulated through: (i) An Ontology Browser that displays a clinical view of the Ontology's hierarchical structure, enabling the user to exploit relations with connected concepts; (ii) Concept Finder that provides on-type suggestions of concepts included in the Ontology and/or MeSH. It can also be used as a translation tool by exploiting multilingual versions of MeSH; (iii) a Query Building tool to construct queries by using concepts from the Ontology and MeSH, as well as, Boolean expression in-between them. Conceptual search retrieves content that does not necessarily contain keywords from the query, actually synonyms and sub-concept are exploited to recall all the document related to a search.
- Free-text search: This is also a conceptual-based search which allows the user to submit queries in natural language, e.g. a complete sentence. The NLP Tool processes the sentence to keep the valuable information and investigates the associations stored in the Knowledge Cache graph database to return the documents that match to the query. Free-text search can also receive a query in a different language than English (currently in Italian, Spanish and Portuguese) and exploit the translated versions of MeSH to build a query with the respective English synonyms.

## 6. Ontology/Thesaurus Module

The purpose of this module is to provide the access to all the domain and linguistic resources required to annotate and retrieve the certified medical reference documents and to enhance the search functionalities therewith.

This module provides access to:

- the COPD, CKD and MLOCC (middle-layer) ontologies [2];
- the MeSH (English) thesaurus and its multilingual lexicographic representations;
- the mapping between MeSH concepts and corresponding Ontology classes;

The CHRONIOUS ontologies encode expert knowledge relevant to the envisaged domain (COPD and CKD) such as to provide a rich science-driven repository of concepts for annotating and searching medical literature in the scope of the project, as well as a wealth of relations to refine domain-specific literature search [10][5].

Furthermore, we have expanded the ontology providing reference for selected terms from MeSH [1]. MeSH is a certified and very rich thesaurus that covers a large part of the concepts pertaining to the medical domains. It is already used to index medical articles, but

- it does not deepen specifically the two diseases considered, so it lacks of the required specificity to support in complex search pertaining COPD and CKD;
- it is not compliant with the newest web standard suggested by the W3C, so it cannot be easily combined with the OWL domain ontologies.

In the project, the former limitations are overcome by the two domain-specific ontologies. However, searching for scientific papers may require to move from concepts strictly related to the aforementioned diseases to concepts less specific and vice versa, thus some connections between the two ontologies and the MeSH concepts are required. For this reason, a MeSH-Ontologies mapping linking the concepts provided by MeSH to the ontologies classes has been provided. To ensure a fruitful connection, we have mapped\encoded the MeSH content into SKOS model following the rules discussed in [11] and serialized the SKOS in RDF. SKOSyfied MeSH thesaurus, its translations and its mapping to concept described in the Diseases Ontologies are accessible by API developed in CHONIOUS which relies on the JENA framework [12] and Ontologies are accessible through standard OWL 2 API [13].

## 7. Ontology Enrichment Tool

The Ontology Enrichment Tool (fig. 2) supports the ontology curator in evaluating new medical documents for the purpose of the ontology maintenance and enrichment in a semi-automatic way.

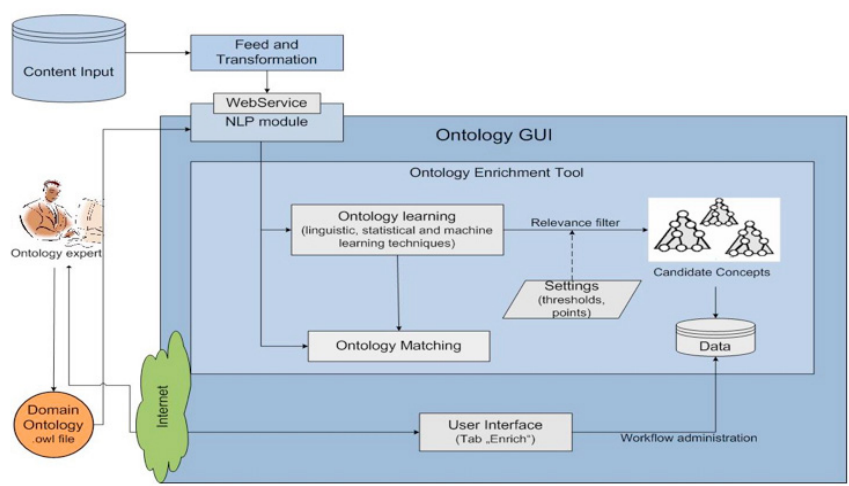

*Figure 2 CHRONIOUS Ontology Enrichment Tool*

In particular, it allows identifying gaps in the ontologies since the enrichment as an iterative process is based on the documents repository itself. Therefore, the tool combines ontology learning techniques and submission functionalities.

### 7.1   Ontology Learning Functionality

Methods for ontology learning are based on linguistic, statistical and machine learning approaches. These techniques are usually combined with each other (as hybrid methods) for concept extraction and taxonomy building. Accordingly the Ontology Enrichment Tool uses the NLP Tool for a linguistic processing of the transformed document data and annotates the text with several features. Based on these annotations candidate concepts will be extracted in a second step.

After the pre-linguistic processing (see Section 4) the extracted candidate concepts are rated based on their relevance, whose estimation is based on the several criteria: the concept's average Term Frequency Inverted Document Frequency (TF.IDF) value with respect to the whole document corpus; matching with a common dictionary, a domain thesaurus and regular expression patterns to determinate the domain relevance of a candidate concept. Also the assignment of synonyms results in a higher relevance, for example when the candidate concept is identified as a synonym of an existing concept (by dictionary synonyms or ontology mapped MeSH thesaurus terms);the relevance of the candidate concept is raised if it is possible to assign a subclass-of relation to it. Such an assignment can be determined by the extraction of vertical relations, linguistic patterns or dictionary hypernyms; co-occurrences of the candidate concept with concepts in sentences can be considered as an indicator of a possible relation extraction; and finally the average distance in the text between the candidate concept and a concept within the corpus is calculated as a benchmark.

## 7.2   Submission Functionality

The GUI of the Ontology Enrichment Tool allows reviewing the extracted data, in particular the list of the candidate concepts and the list of identified concepts. Furthermore the user has the possibility to administrate the candidate concepts in terms of a workflow state ("new", "to validate", "postponed", "accepted" or "rejected"). In this way the submission function of the Enrichment Tool provides suggestions for new concepts (possibly with subclass-of relation assignment) or new labels of existing concepts, but modifications of the ontology must still be done by the ontology expert with an external tool (e.g. Protégé).

In addition, the Enrichment Tool can send emails to registered users in order to remind about certain events, ontologists will be informed if new candidate concepts are available, and Clinicians with ontologies responsibility will be notified, if candidate concepts have to be validated.

## 8.   Evaluation and Comparison with Related Systems

The CHRONIOUS Search Module enables healthcare professionals to access well focused and up to date healthcare information in the form of scientific literature, guidelines, hospital-specific documentation and grey literature. The proposed search mechanism overcome well known search engines like Google Scholar, PubMed and GoPubMed by combining the MeSH potentialities with the two specific domain ontologies. Google Scholar supports the search of documents a pure syntactical search of keywords in the documents, without taking into account any semantic interpretation of the proposed query. Thus, two different queries with equivalent semantics yield different search results. PubMed improves Google Scholar by exploiting the synonymous provided by MeSH, where the thesaurus is used to produce a semantic interpretation of the query keyword mechanism. The GoPubMed search engine, improves PubMed by considering the concepts of Gene ontology (GO) during the query refinement step. In this case, the clinicians are facilitated during the query composition by exploiting both the terms of MeSH and the terms carried by the Gene ontology. Neither PubMed nor GoPubMed are supported by the specific disease knowledge provided by COPD and CKD ontologies. This additional representational power is exploited to improve the indexing of medical literature. An accurate evaluation of the two domain ontologies and a quantitative comparison between the CHRONIOUS search engine and Pubmed have been provided.

The CHRONIOUS ontologies and the mapping relations between MeSH concepts and ontology classes have been validated by medical experts. Questionnaires measuring the agreement among medical experts have been completed by COPD and CKD experts. The evaluation result regarding the correctness of the developed ontologies shows that most of the ontology classes have been defined correctly, a few ontology classes need to be modified (eight of 964 COPD classes and nine of 972 CKD classes), and together 150 new classes should be complemented additionally (80 new classes for COPD ontology and 70 for CKD ontology). That is less than 7.8% of the total number of defined ontology classes. According to the feedback of medical experts to the correctness of MeSH mapping relations, 42 of 120 mapping relations between MeSH concepts and COPD ontology classes, as well as 43 of 138 mapping relations between MeSH concepts and CKD ontology classes have been modified. Besides the correctness evaluation of CHRONIOUS ontologies and MeSH mapping, search performance of the CHRONIOUS Conceptual Search, i.e., the exactness and completeness of its search result have been evaluated as well. Exactness (or Precision) measures the number of correctly found documents – documents which are relevant to the search query – as a percentage of the number of documents found, whereas completeness (or Recall) measures the number of correctly found documents as a percentage of the total number of correct documents. Usually, Precision and Recall scores are not discussed in isolation. Instead, both are combined into a single measure – the F-measure [15], which is the weighted harmonic mean value of Precision and Recall.

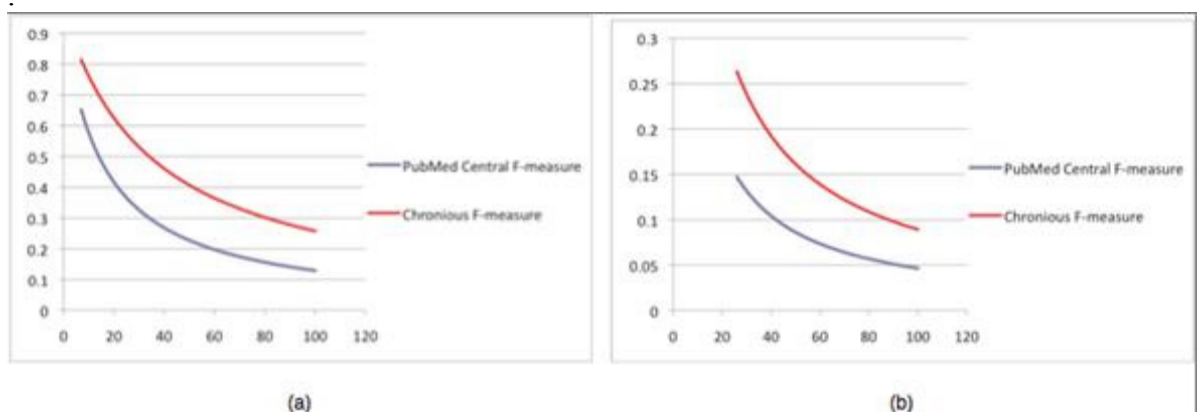

*Figure 3: F-measure comparison between CHRONIOUS Conceptual Search and PubMed Central with search query (a) "Inhaler Device" and (b) "PostBronchodilator Spirometry". The X-axis presents the estimated total number of existing relevant documents in the repository for the corresponding search query (increased incrementally).*

To measure the search performance of CHRONIOUS Conceptual Search, seven commonly used COPD terms and five CKD terms have been defined by medical experts as search queries. The F-measure of search results of each search query with the CHRONIOUS Conceptual Search has been calculated and compared with one of the most popular text-based search engines – PubMed Central. The test result shows that the overall search performance of CHRONIOUS Conceptual Search with defined search queries is in most cases better than PubMed Central. **Error! Reference source not found.** shows the F-measure comparison results with two search terms: "Inhaler Device" and "PostBronchodilator Spirometry"[1]. Furthermore, the overall end user satisfaction of the CHRONIOUS Search Module has been investigated with the medical experts by a questionnaire. The evaluation has shown that 90% of them generally accept the Conceptual Search development and 60% of them gave positive evaluation to the Free-text Search option.

---

[1] The comparison of search performance between PubMed Central and CHRONIOUS Search Module needs to be executed with same document database. In the COPD case, 930 documents which were published from January 1st to December 31st, 2008 and contained the term "COPD" in their text (incl. title, abstract, text body, figure/table caption, etc.) have been downloaded from PubMed Central and uploaded into the CHRONIOUS document repository. The search results presented in ***Error! Reference source not found.*** have been retrieved among this set of documents.

 

## 9. Conclusions and Summary Recommendations (Lessons Learnt)

In conclusion the CHRONIOUS search module aims at complementing out of shell search modules such as Google Scholar, PubMed and GoPubmed not only by improving the search capabilities within the specific pathologies, but also by providing the possibility to index and retrieve hospital-specific internal documentation. The ontology-thesaurus approach can overcome limitations of pure syntactical document search, in which two queries having different syntax but equivalent semantics yield different search results, and it can ease multilingual search issues. In addition, the adopted multi-layer ontology design offers a methodological support to separate concepts shared among different diseases from those diseases specific, that turns out in disease-specific ontology based on well founded and formalised concepts, and a knowledge architecture where further disease ontologies can be plugged-in. The development of the prototype has demonstrated that the currently available open source software provide sufficient functionalities for realising ontology-based literature retrieval systems and ontology enrichment capabilities. Despite of this, ontology enrichment cannot be either fully automatic or directly used by clinicians: both clinicians and ontology engineers are needed to semantically validate and to correctly insert and correlate the candidate concepts. In fact, from our experience, clinicians have difficulties in browsing formal ontologies. For this reason, it is necessary to provide clinician suitable views of the ontologies and tools for a user friendly concepts' selection.

## Acknowledgment

This research has been performed within the project CHRONIOUS (EU Contract N. FP7-ICT-2007–1–216461). The authors want to thanks all the partners for their support. This paper was written as a joint contribution; authors are listed in alphabetical order grouped by their affiliations, besides they have equally contributed to this paper.